# Spintronics in semiconductors

Z. Wilamowski[1] and A. M. Werpachowska[1,2]

[1]Institute of Physics PAS, 02-668 Warsaw, Poland

[2]College of Science by the Polish Academy of Sciences, Warsaw, Poland

**Abstract**

For the last years spin effects in semiconductors have been of great interest not only in the context of solid state physics, but also for their potential usage in technology. In this paper we give a short review of spintronic materials, in which electron spin as an additional degree of freedom is exploited. Afterwards, we discuss the properties of classic, non-magnetic semiconductors, where the efforts are put on enriching the traditional semiconductor technology engaging the electrical effects of spin effects. Various phenomena and scientific state of the art is highlighted.

## 1. Introduction

Next to the electron charge, the electron spin corresponds to an additional degree of freedom, which could be used for information storage and processing. To control electron and spin states, one has to measure different physical quantities. Electrical properties are characterized by electrical conductivity, carrier mobility, voltage profile and electrical current, while spin properties are characterized by magnetization, magnetic resonance frequencies and spin relaxation rates. There are also different tools that can be used to manipulate electron charge and spin state. Electronic devices are controlled predominantly by applied voltages, while for the manipulation of spin state one has to use magnetic field. In contrast to the voltages, the magnetic field cannot be applied locally. For those reasons spintronics needs sophisticated solutions for the various classes of spintronic materials. In this paper we discuss the properties of novel spintronic materials and classic, non-magnetic semiconductors and a possibility of their use for electronic devices.

## 2. Magnetic materials

Ferromagnetic metals own their great career as spintronic materials to the spin effects they display. The most popular in practical applications are the giant magnetoresistance and tunneling magnetoresistance in devices built of ferromagnetic metals. The resistance of layered structures depends on the mutual orientations of magnetization in the neighbouring layers. Such elements are already commonly applied as reader heads for magnetic memories. The weak point of the ferromagnetic metals is that it is hard to modify their properties with an applied electrical field.

The possibility of changing the physical properties with an applied electrical field is the feature of semiconducting materials. The materials of the greatest interest to technology are Diluted Magnetic Semiconductors (DMS). To make a typical DMS, like (Cd,Mn)Te, we need to substitute a part of diamagnetic atoms in a classic semiconductor with atoms of the transition metal [1]. Such highly diluted material is paramagnetic. DMS acts in the absence of an external field like a semiconductor. Application of the external field results in the strong spin splitting of the conduction band and the valence band. The origin of the spin splitting is the exchange coupling sp-d between delocalized carriers and core spins.

The most characteristic feature of DMS is giant spin splitting, which increases the spin polarization of the carriers, usually present in small number. It is easy to achieve 100% polarization in laboratory conditions. Since the onset of spintronics, there were attempts to use this property of the DMS to build spintronic elements. The first idea was to build a Giant Magnetic Resistance (GMR) structure. The attempts to build a hybrid ferromagnetic metal-semiconductor-ferromagnetic metal structure failed, due to the fact that a small semiconductor conductance suppresses the spin current [2] but enhances the spin accumulation.

However, efforts to inject spins from semi-magnetic semiconductors were much more successful. As Fiederling at al. [3] have shown, the injection efficiency from DMS to GaAs is near 100%. Spin polarization was estimated from the circular polarization ratio from GaAs. It is also possible to build spin transistors [4] based on DMS.

Increasing of the carrier concentration in DMS may lead to the appearance of ferromagnetic phase. Given sufficiently large band filling, not only do the local spins cause a spin polarization of the carriers, but conversely, polarized carrier band may, through the same p-d exchange, cause the local spins' polarization. Munekata and Ohno [5-8] gave the first evidence of ferromagnetism in (In,Mn)As and (Ga,Mn)As materials. The first theoretical description should be attributed to Dietl [6].

(Ga,Mn)As is a canonical semiconductor. The critical temperature of the currently manufactured layers exceeds 170 K, and the alloy retains all semiconducting properties. The electrical control of ferromagnetism is possible, in particular, the critical temperature can be changed by applying an electrical field [6,7]. (Ga,Mn)As is also a very good source of polarized electrons, which allows to inject spins to normal semiconductors [8]. Because of the spontaneous magnetization in the ferromagnet, it is not necessary to apply an external magnetic field in order to achieve a stream of spin polarized carriers. The structure similar to TMR structures built on (Ga,Mn)As shows a very large magnetoresistance [9] and tunneling anisotropic magnetoresistance [10]. In the tunneling transport regime the magnetoresitance exceeds 2000%. Yamanouchi et al. demonstrated the current control domain wall motion [11]. Astakhov et al. [12] demonstrate the spin switching between two metastable magnetization states in materials which posses uniaxial anisotropy. It may be induced by applying the magnetic field, but also by a laser pulse [13,14].

### 3. Spin in classic semiconductors

In classic metals and semiconductors magnetism plays a secondary role. Because of the Pauli principle, the equal filling of the up and down spin subbands leads to the cancellation of magnetic momenta. In external magnetic field a weak carrier magnetization appears, but the Pauli susceptibility is very low. In most metals and semiconductors the dependence of electrical properties on spin properties is negligible. The connection between electrical and magnetic properties is visible only in exceptional cases. One of them is the dependence of the resistance of two-dimensional electron gas on its spin polarization. The only mechanism linking electrical and magnetic properties is the existence of a spin-orbit coupling, which leads to the spin splitting of the bands.

The zero field splitting of spin subbands occurs only in semiconductor structures with sufficiently low symmetry. In general, one distinguishes between the Dresselhaus field [15] and the Rashba field [16,17]. Spin splitting is equivalent to an effective spin-orbit field. The Dresselhaus field is the consequence of the lack of the crystal inversion symmetry. It occurs, e.g., in zinc blend structure, i.e., in all II-VI and III-V semiconductor compounds but it is forbidden by symmetry conditions in bulk silicon. The Bychkov-Rashba field is a consequence of the lack of the mirror symmetry in 2D structures. It occurs also in silicon 2D structures. The direction of the spin-orbit field, $H_{so}$, is perpendicular to carrier $k$-vector. In the case of Bychkov-Rashba field it is in-plane oriented. Generally the inversion of the $k$-vector direction leads to inversion of $H_{so}$. As a consequence, in thermal equilibrium the sum of all spin-orbit fields acting on the carrier system vanishes. It follows that most of the spintronic (dependent on both magnetic and electrical properties) effects vanish in the thermal equilibrium, too. Therefore, we need to search for phenomena engaging both the spin and electron properties only in the systems in thermodynamic nonequilibrium. Below we shall review the class of spin effects induced by electron current, namely spin Hall, spin manipulation by electrical current, and the class of spin fotovoltaic effects.

When an electrical current is applied and the Fermi sphere is moved from the center, the total spin-orbit field acting on the ensemble of the electron spins is not anymore zero. The electron spins begin the precession and a tendency to create additional spin polarization appears. Both effects are diminished by spin relaxation. However, when the frequency of RF current is much higher than the spin relaxation rate, the spin polarization is not affected by electrical current and the effective RF magnetic field is the main consequence of the current. As it was demonstrated by Wilamowski and Jantsch [18], such SO field can play the role of microwave magnetic field leading to an additional resonance absorption. As a consequence three different ESR signals are observed in 2D electron gas: a classic absorption caused by the magnetic component of microwave field, SO field caused by the electrical current and an additional, so-called polarization signal, resulting from the change of electrical conductivity under resonance condition. This effect reflects the dependence of the conductivity of the high mobility 2D electrons on spin polarization and allows for electrical measurement of spin structure.

When DC or low frequency current is applied to the semiconducting layered structure, two additional processes have to be considered: the possible precession of spin polarization around the total magnetic field, including the effective SO field, and a spin relaxation which leads to an additional spin polarization caused by the SO field. The interference of these effect leads to occurrence of spin Hall effect [19, 20, 21, 22]. The spin accumulation at the edges of non-magnetic semiconducting samples has been demonstrated experimentally [19, 20]. The effect can occur also in the absence of any external magnetic field, reflecting the intrinsic character and indicating the SO field as the responsible for the effect.

Because in 2D electron gas the SO field is only in-plane oriented, the resulting magnetization is also expected to be in plane oriented. Moreover, the spin polarization of individual electron is parallel to an individual SO field. Thus such SO field does not cause any precession which would lead to a perpendicular component of the carriers magnetization, which seems to be a necessary condition for the Hall effect. For that reasons, the modeling of spin Hall effect requires a simultaneous discussion as well of momentum scattering, which allows to tilt the SO field and spin directions, as of spin precession and spin relaxation which could lead to the occurrence of the perpendicular component of carrier magnetization [21, 22]

It is well known that circularly polarized light, due to spin dependent transition probabilities, causes the spin polarization of excitons and of carrier spins. The occurrence of the spin polarization means that up and down spin subbands become differently occupied. Consequently, the Fermi $k$-vectors for up and down spins are different. For symmetry reasons, however, the spin polarization only does not lead to any macroscopic current, but the symmetry changes when an external magnetic field is additionally applied. As it is shown by Ganichev et al. [23, 24], the sample illumination leads to the occurrence of electrical current. Such spin galvanic effect (Hanle effect) is ruled by a velocity and spin dependent momentum relaxation and by a precession of magnetization in the external magnetic field. The complex direction dependence of the Dresselhaus SO field result in the complex dependence of the spin galvanic effect on the direction of light illumination and the magnetic field versus crystallographic directions [24].

Also, the reversed galvanic effect has been observed. In that case the electrical current causes the non-equilibrium spin polarization. The idea of such effect has been proposed by Edelstein 15 years ago [25]. The experimental evidence has been found in 2004 by Silov et al. [26], who showed that for the proper experimental geometry the electrical current results in circular photoluminescence. Kato el al. [27] also established the current induced spin polarization predictions by the measurement of Faraday rotation.

## 5. Summary

To summarise, in the present days the fast development of spintronics shows that spintronic elements can be built not only of magnetic materials, such as ferromagnetic metals or ferromagnetic semiconductors. A wide spectrum of spintronic effects can also be found in the classic, non-magnetic semiconductors.

## 6. Acknowledgement

This work supported by PBZ-KBN-044/P03/2001.